\documentclass{article}

\usepackage{microtype}
\usepackage{graphicx}
\usepackage{subfigure}
\usepackage{booktabs}
\usepackage{hyperref}

\usepackage[accepted]{icml2024}

\usepackage{amsmath}
\usepackage{amssymb}
\usepackage{mathtools}
\usepackage{amsthm}

\DeclareMathOperator{\sigmoid}{sigmoid}

\usepackage[capitalize,noabbrev]{cleveref}

\theoremstyle{plain}

\theoremstyle{definition}

\theoremstyle{remark}

\usepackage{array}
\usepackage{multicol}

\icmltitlerunning{Toward Fully Self-Supervised Multi-Pitch Estimation}

\begin{document}

\twocolumn[
\icmltitle{Toward Fully Self-Supervised Multi-Pitch Estimation}

\begin{icmlauthorlist}
\icmlauthor{Frank Cwitkowitz}{ur}
\icmlauthor{Zhiyao Duan}{ur}
\end{icmlauthorlist}

\icmlaffiliation{ur}{Department of Electrical \& Computer Engineering, University of Rochester, Rochester, New York, USA}

\icmlcorrespondingauthor{Frank Cwitkowitz}{fcwitkow@ur.rochester.edu}

\icmlkeywords{Multi-Pitch Estimation, Self-Supervised Learning, Music Transcription, Timbre-Invariance, Geometric-Equivariance}

\vskip 0.3in
]

\printAffiliationsAndNotice{}

\begin{abstract}
Multi-pitch estimation is a decades-long research problem involving the detection of pitch activity associated with concurrent musical events within multi-instrument mixtures.
Supervised learning techniques have demonstrated solid performance on more narrow characterizations of the task, but suffer from limitations concerning the shortage of large-scale and diverse polyphonic music datasets with multi-pitch annotations.
We present a suite of self-supervised learning objectives for multi-pitch estimation, which encourage the concentration of support around harmonics, invariance to timbral transformations, and equivariance to geometric transformations.
These objectives are sufficient to train an entirely convolutional autoencoder to produce multi-pitch salience-grams directly, without any fine-tuning.
Despite training exclusively on a collection of synthetic single-note audio samples, our fully self-supervised framework generalizes to polyphonic music mixtures, and achieves performance comparable to supervised models trained on conventional multi-pitch datasets.
\end{abstract}

\section{Introduction}\label{sec:introduction}
Music is comprised of auditory events generated by exciting the strings, chambers, or surfaces of an instrument.
Pitched events or notes are characterized by air particle oscillations, each with varying strength and frequency at an integer multiple (in Hz) of a fundamental frequency (F0) \cite{muller2015fundamentals}.
A note's F0 directly informs its pitch and captures articulations such as vibrato, while the presence and relative strength of the remaining oscillations or harmonics influence its unique timbral properties \cite{klapuri2006signal}.
Multi-pitch estimation (MPE) seeks to detect all F0 activity across regular time steps within polyphonic audio signals.
This is an important task with broad interest in Music Information Retrieval (MIR) research and exciting applications in music analysis\footnote{\url{https://www.sonicvisualiser.org}},  composition\footnote{\url{https://scorecloud.com}},  editing\footnote{\url{https://www.celemony.com/en/melodyne}}, education\footnote{\url{https://yousician.com}}, indexing and search\footnote{\url{https://www.deezer.com/explore/en-us/features/songcatcher/}}, etc.
Pitch is foundational for various musical concepts surrounding melody and harmony.
As such, MPE is also critical for several downstream MIR tasks, including automatic music transcription (AMT) \cite{benetos2018automatic} and automatic chord recognition (ACR) \cite{pauwels201920}.
Furthermore, MPE has the potential to inform systems addressing other challenging MIR tasks like source separation \cite{schulze2023unsupervised}, polyphonic audio synthesis \cite{renault2023ddsp}, or bandwidth extension \cite{grumiaux2023efficient}.

\begin{figure}[t]
\centering
\includegraphics[width=0.98\columnwidth]{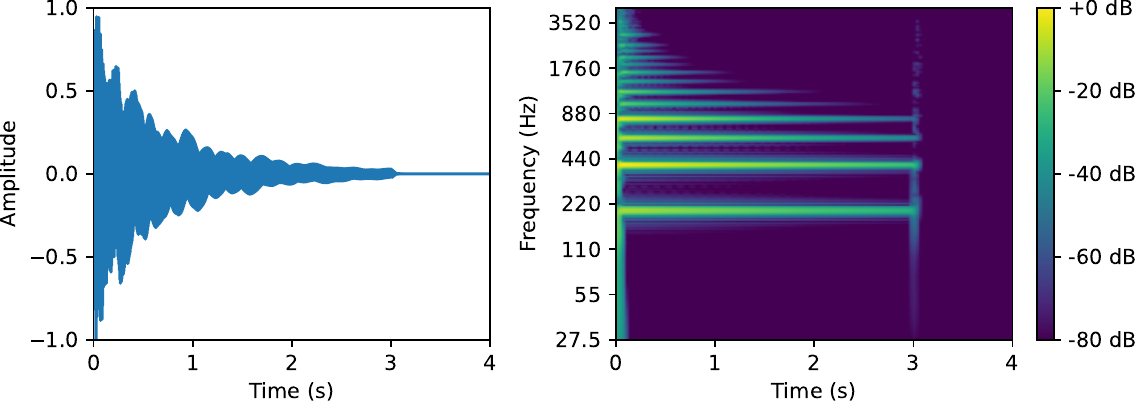}
\caption{Example note signal (left) and its corresponding power spectrogram (right) from the NSynth dataset \cite{engel2017neural}, which is used to train the proposed self-supervised framework.}
\label{fig:sample}
\end{figure}

Although monophonic or single-pitch estimation systems have reached near-perfect accuracy \cite{morrison2023cross}, the polyphonic or multi-pitch case has proven to be much more difficult.
This is largely due to the problem of overlapping harmonics and ambiguity regarding the number of simultaneously active notes at a given time \cite{klapuri1998number}.
However, this difficulty is further compounded by the lack of availability of large-scale comprehensive datasets for training MPE systems.
At present, acquiring multi-pitch annotations essentially consists of running monophonic pitch trackers on isolated stems from mixtures of monophonic instruments and performing manual error-correction \cite{duan2010multiple, bittner2014medleydb, li2018creating}.
Alternatively, to avoid manual correction without propagating errors, audio stems can be re-synthesized directly from the pitch estimates \cite{salamon2017analysis}.
These procedures have significant limitations in terms of scalability and the types of musical instruments, passages, and environments that can be supported.

Additionally, supervision has never been the primary learning paradigm for humans to learn to transcribe simultaneous pitches.
Humans, especially people without absolute pitch, rely on the intrinsic patterns of auditory representations and their relations across pitch to transcribe music \cite{bregman1994auditory}, requiring only one or a few reference pitches (\textit{i.e.}, labels).
This more closely resembles the self-supervised learning (SSL) paradigm in machine learning, which has been garnering increasing attention due to its profound success in learning rich representations for complex data spanning various modalities \cite{ericsson2022self}.
SSL has already been successfully applied to monophonic pitch estimation, by leveraging the pitch-transposition equivariance property of the Constant-Q Transform (CQT) \cite{gfeller2020spice, riou2023pesto}, a common representation for MIR, or using model-based approaches \cite{engel2020self}.
However, MPE is a much more complicated problem, and no attempts have yet been made to adopt an SSL approach for the task.

In this work, we present a fully self-supervised framework for MPE (SS-MPE), consisting of a Harmonic Constant-Q Transform (HCQT) \cite{bittner2017deep} feature extraction stage, a fully convolutional 2D autoencoder \cite{cwitkowitz2024timbre}, and a closed-loop suite of self-supervised training objectives.
Our objectives collectively encourage the model to produce multi-pitch salience-gram estimates directly, bypassing the need for fine-tuning or annotated data.
The first three objectives encourage concentration of energy around strong F0 candidates, derived via a simple weighted harmonic average \cite{klapuri2006multiple}.
The remaining two objectives promote invariance to timbre transformations and equivariance to geometric transformations, intrinsic properties of pitch and the CQT, respectively.
Our model is trained entirely on the NSynth dataset \cite{engel2017neural}, which comprises 4-second monophonic note samples (see Figure \ref{fig:sample}) synthesized from commercial virtual instruments.
While a global pitch label is available for the samples in NSynth, our framework is completely agnostic to them.

Remarkably, without training on mixtures of the monophonic samples or performing any sort of music language modeling, SS-MPE is able to generalize to polyphonic music and yields MPE performance approaching that of supervised methods, requiring only a simple peak-picking and thresholding procedure at the output.
This result is very promising and merits further investigation into SSL for MPE and related tasks.
Our work is completely open-source\footnote{All code is publicly available at \url{https://github.com/cwitkowitz/ss-mpe}.}, and can be reproduced with minimal computational resources.
This paper seeks only to establish a foundation for applying SSL techniques to MPE. We suggest several directions for future work, including scaling the approach to large polyphonic music datasets and designing self-supervised objectives to exploit additional properties of music and audio.

\section{Related Work}\label{sec:related_work}
MPE for both speech and music signals has been an active area of research for several decades \cite{decheveigne2006multiple, christensen2009multi}.
Traditional methods are largely based around signal processing techniques and statistical modeling, including spectral analysis \cite{klapuri2003multiple, yeh2009multiple, zhou2009computationally}, auditory modeling \cite{klapuri2008multipitch}, maximum \textit{a posteriori} \cite{kameoka2007multipitch} or maximum likelihood \cite{emiya2009multipitch, duan2010multiple} estimation, and filtering or subspace techniques \cite{christensen2008multi, christensen2011joint}.
Another class of methods perform factorization on spectral features \cite{bertin2010enforcing, benetos2015efficient} or the audio waveform \cite{cogliati2017piano} to estimate note templates and their corresponding activations.
While traditional methods are generally interpretable and yield reasonable performance, they often incorporate several tunable parameters and fail to generalize to the complexity of arbitrary real-world music \cite{benetos2013automatic}.
Moreover, most research has investigated instrument or domain-specific characterizations of MPE in order to simplify the problem. These approaches tend to be targeted toward piano transcription, mainly due to the relative convenience of collecting corresponding note annotations for audio from MIDI-driven acoustic pianos (\textit{e.g.}, Disklavier) \cite{emiya2009multipitch, hawthorne2018enabling}.

Another related task is AMT, where the aim is to directly estimate note events or pitches quantized to a musical scale \cite{benetos2018automatic}.
Typically, AMT systems either perform MPE implicitly \cite{gardner2021mt3} or consist of a frame-level MPE module followed by post-processing to produce note-level predictions \cite{benetos2011joint, elowsson2020polyphonic}.
The methodologies and challenges for MPE and AMT are nearly identical, and their terminology is frequently used interchangeably in the literature.

With the explosion of deep learning, various convolutional-recurrent neural networks were subsequently proposed for piano transcription \cite{bock2012polyphonic, sigtia2016end, hawthorne2018onsets} and other instrument-specific AMT tasks.
These models can generally be factored into an acoustic component which extracts rich features from input spectra and a music language component which aggregates the features across time before making final predictions.
Currently, sequence modeling \cite{yan2021skipping} and transformer-style architectures \cite{hawthorne2021sequence, toyama2023automatic} dominate piano transcription, with systems nearing superhuman performance.
In parallel, there has also been work to build more general MPE systems \cite{bittner2017deep} based on deep learning and to reuse insights from piano transcription research for multi-instrument AMT \cite{bittner2022lightweight}.
In particular, transformer-style architectures applied to this domain have demonstrated remarkable efficacy \cite{gardner2021mt3, lu2023multitrack}.
U-Net architectures have been also applied to AMT as frontends to learn richer feature representations \cite{pedersoli2020improving, weiss2022comparing}, as well as to predict multi-pitch activations directly \cite{wu2020multi}.
These autoencoders transform input spectrograms into latent codes with reduced dimensionality, which are then decoded into an output representation matching the original input size.
This type of architecture is especially notable for MPE, given the typically high correlation between spectrograms and multi-pitch salience-grams, as well as their matching dimensionality.

While these methods have undoubtedly advanced research on MPE, they are still fundamentally limited by the availability of large-scale and diverse training data.
As such, alternative methods have been proposed to extract more out of the data that is currently available.
Some works leverage multi-task learning \cite{manilow2020simultaneous, tanaka2020multi, hung2021transcription, cheuk2022jointist} to simultaneously separate and transcribe various sources, while others employ auxiliary reconstruction objectives \cite{cheuk2021effect, cheuk2021reconvat, wu2023mfae, cwitkowitz2024timbre} to promote a more thorough modeling of music signals.
Additionally, several methods have utilized weak supervision in order to train models with weakly-aligned data \cite{weiss2021learning, maman2022unaligned, krause2023soft, krause2023weakly}.
Pre-training methodologies have also been proposed to learn from random mixtures of pre-transcribed monophonic audio in-the-wild \cite{simon2022scaling} or pitch annotations without corresponding audio \cite{cheuk2023diffroll}.
Finally, many works have leveraged synthetic datasets \cite{engel2017neural, manilow2019cutting, ostermann2023aam}, however it is generally recognized that there exists a significant domain gap between synthesized and real-world audio.

Following the fields of natural language processing and computer vision, there has also recently been an increasing focus on designing self-supervised techniques to learn robust feature representations for downstream tasks from large databases of unlabeled audio.
One such technique is to leverage training objectives based on masked reconstruction in order to teach models the important characteristics and structure of audio \cite{tagliasacchi2020pre} and music \cite{li2023mert} signals.
Another is to use a contrastive learning objective \cite{chen2020simple}, which maximizes the similarity between the learned representations for randomized label-preserving transformations of a unique training example, while simultaneously maximizing the dissimilarity between the learned representations for disparate samples.
Contrastive learning has been applied to general audio \cite{al2021clar, saeed2021contrastive} and MIR \cite{spijkervet2021contrastive, mccallum2022supervised, mccallum2024effect} in order to teach models how to organize information about signals that will be relevant for downstream processing.
SSL has also been applied directly to specific MIR tasks including monophonic pitch estimation \cite{gfeller2020spice, riou2023pesto}, singer identity representation learning \cite{torres2023singer}, beat tracking \cite{desblancs2023zero}, and tempo estimation \cite{henkel2024tempo}.
For pitch estimation, the pitch-transposition equivariance property of the input CQT can be exploited to learn relative differences in pitch.
Self-supervised and unsupervised techniques based on the principles of differentiable digital signal processing (DDSP) \cite{engel2020ddsp} have also been proposed for monophonic pitch estimation \cite{engel2020self, hayes2023sinusoidal, torres2024unsupervised}.
Finally, multi-speaker self-supervised pitch estimation based on physical modeling of the human vocal tract has been carried out \cite{li2022multi}, though this approach is explicitly designed for two speakers.

\section{Method}\label{sec:method}
In this section, we detail our proposed framework SS-MPE.
We begin by discussing our HCQT feature extraction module, which provides strong \textit{a priori} information for MPE and has several important implications for the design of our self-supervised objectives.
We then outline the autoencoder network employed within our framework and the process of obtaining F0 estimates from the output multi-pitch salience-grams.
Finally, we present our SSL objectives, which concentrate energy around strong F0 candidates and exploit the properties of timbre-invariance and geometric equivariance.

\subsection{Feature Extraction}\label{sec:feature_extraction}
\begin{figure}[t]
\centering
\includegraphics[width=0.98\columnwidth]{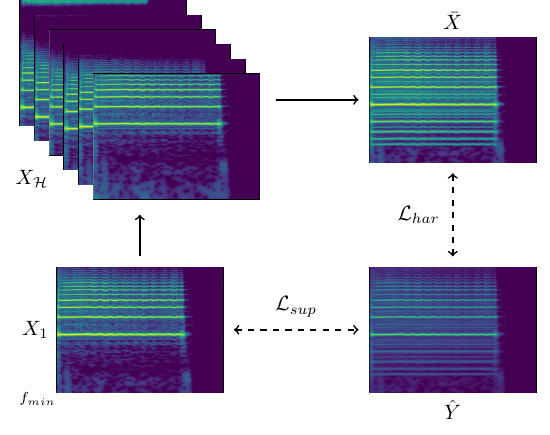}
\caption{HCQT power spectrogram (in dB) for an audio signal (top-left), the corresponding first-harmonic channel (bottom-left) and weighted harmonic average (top-right), and an example multi-pitch salience-gram estimate (bottom-right). Self-supervised objectives ($\mathcal{L}_{har}$ and $\mathcal{L}_{sup}$) are employed to encourage the model to concentrate energy around strong F0 candidates.}
\label{fig:hcqt}
\end{figure}

The CQT is a time-frequency representation derived from basis functions with geometrically spaced center frequencies and a constant Q-factor \cite{brown1991calculation}.
It has been applied widely across various MIR tasks to compute the frequency response for frequencies aligned with the Western music scale \cite{schorkhuber2010constant}.
Illustrations of the CQT spectrogram are provided in Figure \ref{fig:sample} (right) and \ref{fig:hcqt} (bottom-left).
Like other time-frequency representations, the CQT is equivariant to time shifting and stretching.
However, the CQT is also pitch-transposition equivariant, meaning that pitch shifting a time-domain signal will correspond to a frequency (vertical) translation in its CQT spectrogram.
These properties have been leveraged to learn shift-invariant representations for relevant MIR tasks \cite{lattner2019learning}, and to design self-supervised training objectives for pitch estimation \cite{gfeller2020spice, riou2023pesto}.

In this work, we utilize the HCQT, a simple extension which stacks multiple CQTs along a third channel axis \cite{bittner2017deep}, illustrated in Figure \ref{fig:hcqt} (top-left).
The individual CQTs $X_h$ are computed relative to the harmonics and subharmonics $h \in \mathcal{H}$ of a reference frequency $f_{min}$.
We follow the originally proposed set $\mathcal{H} = \{0.5, 1, 2, 3, 4, 5\}$.
Although we are only interested in detecting F0s from the first harmonic channel ($h = 1$), the inclusion of prime-number harmonic channels ($h \in \{2, 3, 5\}$) can offer theoretical advantages \cite{klapuri1998number}, and the sub-harmonic ($h = 0.5$) and octave ($h \in \{2, 4\}$) channels can help to avoid octave errors.
Since it is simply a collection of CQTs, the HCQT maintains the aforementioned equivariance properties.

Apart from providing richer spectral features, the HCQT allows neural networks, in particular those based on convolution, to leverage the channel dimension in order to more easily discover harmonic relationships for MIR tasks \cite{bittner2017deep}.
Methodologies based on leveraging convolutional kernels with harmonic structure \cite{zhang2019deep, wang2020harmonic, wang2020enhancing, wei2022harmof0, wei2022hppnet} have also been proposed to exploit these relationships within a single-channel CQT spectrogram.
However, many harmonics do not perfectly align with the center frequencies of single-channel spectrogram bins. The HCQT covers these frequencies explicitly by computing a separate CQT for each harmonic.

We utilize a convolution-based implementation \cite{cwitkowitz2019end} to compute decibel-scale ($[-80, 0]$ dB) HCQT\footnote{In practice, the Variable-Q Transform (VQT) \cite{schorkhuber2014matlab} is computed for each harmonic in order to increase computational efficiency and time resolution at lower frequencies.} power spectrograms with a $-80$ dB cutoff for each audio signal.
These features are re-scaled and expressed as $X_\mathcal{H} \in [0, 1]^{6 \times K \times N}$, where $K = 440$ bins with $b_{oct} = 60$ bins per octave ($5$ bins per semitone) and $N$ is the variable frame-length which depends on the input audio size.
The reference frequency for the first-harmonic channel is $f_{min} = 27.5$ Hz, which corresponds to \texttt{A0}, the lowest note on a standard piano.
All audio is resampled to $22,050$ Hz before processing, and we use a hop size of $256$ samples or approximately $11.6$ ms to compute the HCQT spectrograms.

\subsection{Model}\label{sec:model}
We adopt the model from the Timbre-Trap low-resource MPE framework \cite{cwitkowitz2024timbre} as the backbone neural network for SS-MPE.
Timbre-Trap employs a fully convolutional 2D autoencoder inspired by the SoundStream audio compression model \cite{zeghidour2021soundstream}.
The encoder comprises four blocks with dilated convolutional layers, residual connections, and strided convolutions.
The decoder mirrors the encoder, but with transposed convolutions in place of strided convolutions.
The base model is largely left unchanged, though we adapt the input and output channel sizes for our particular usage and remove the latent channel mechanism used in the original paper to switch between reconstruction and transcription mode.
We also insert layer normalization after each block of the encoder and decoder for improved training stability.
In total, the model contains roughly 545K trainable parameters.

Our framework leverages this model to process input HCQT spectrograms $X_{\mathcal{H}}$ by downsampling them along the frequency axis and projecting them into increasing channels, ultimately producing $N$ latent vectors of size $128$.
The latent vectors are subsequently upsampled to the original dimensionality and aggregated into a single channel of logits.
Through our self-supervised objectives, the model implicitly learns to produce multi-pitch salience logits, which are interpreted as multi-pitch salience-grams $\hat{Y} \in [0, 1]^{K \times N}$ by applying $\sigmoid$ activation.
During inference, final F0 estimates are obtained for each frame simply by performing local peak-picking across the frequency axis and applying an un-tuned threshold of $0.5$.

\subsection{Energy Concentration}\label{sec:energy_concentration}
Our first set of self-supervised objectives is designed to influence the model to concentrate energy around strong F0 candidates by leveraging the strong \textit{a priori} information encoded in the input HCQT spectrogram and to adopt an implicit bias toward sparsity.
We first compute a weighted average across the linear-scale HCQT power spectrogram
\begin{equation}
\label{eq:harmonic_average}
\bar{X}^{(lin)} = \sum^{5}_{h = 1} \frac{1}{h^4} X_{h}^{(lin)},
\end{equation}
excluding the sub-harmonic and loosely following harmonic weighting (12 dB decay per octave) precedent \cite{klapuri2006multiple}. We then convert linear-scale $\bar{X}^{(lin)}$ back to decibel scale $\bar{X}$
and formulate two separate losses, the \textit{harmonic} loss $\mathcal{L}_{har}$ and \textit{support} loss $\mathcal{L}_{sup}$. These serve as the positive-only and negative-only components of a continuous binary cross-entropy (BCE) loss, respectively. The harmonic loss
\begin{equation}
\label{eq:harmonic_loss}
\mathcal{L}_{har} = - \frac{1}{N} \sum^{N - 1}_{n = 0} \sum^{K - 1}_{k = 0}
\bar{X}[k, n] \log\left(\hat{Y}[k, n] \right)
\end{equation}
is computed with respect to the positive activations of the weighted harmonic average of the HCQT, and encourages the model to concentrate energy in output $\hat{Y}$ around frequencies with strong energy in $\bar{X}$, \textit{i.e.}, frequencies that manifest strong harmonics overall. However, $\bar{X}$ can also contain energy at frequencies that do not manifest strong energy in $X_1$. Such frequencies cannot correspond to F0s, except for in rare cases where there is a missing fundamental. Therefore, the support loss
\begin{equation}
\label{eq:support_loss}
\mathcal{L}_{sup} = - \frac{1}{N} \sum^{N - 1}_{n = 0} \sum^{K - 1}_{k = 0}
\left(1 - X_1[k, n]\right) \log\left( 1 - \hat{Y}[k, n] \right)
\end{equation}
is computed with respect to the complement of activations in $X_1$ to mitigate this issue. Finally, we introduce an $L_1$-norm \textit{sparsity} loss
\begin{equation}
\label{eq:sparsity_loss}
\mathcal{L}_{spr} = \frac{1}{N} \sum^{N - 1}_{n = 0} \sum^{K - 1}_{k = 0}
\left| \hat{Y}[k, n] \right|,
\end{equation}
which ideally reduces spurious activations and encourages the output of the model to utilize less frequency bins overall.
An example multi-pitch salience-gram produced using these three objectives is presented in Figure \ref{fig:hcqt} (bottom-right).

\begin{figure}[t]
\centering
\includegraphics[width=0.98\columnwidth]{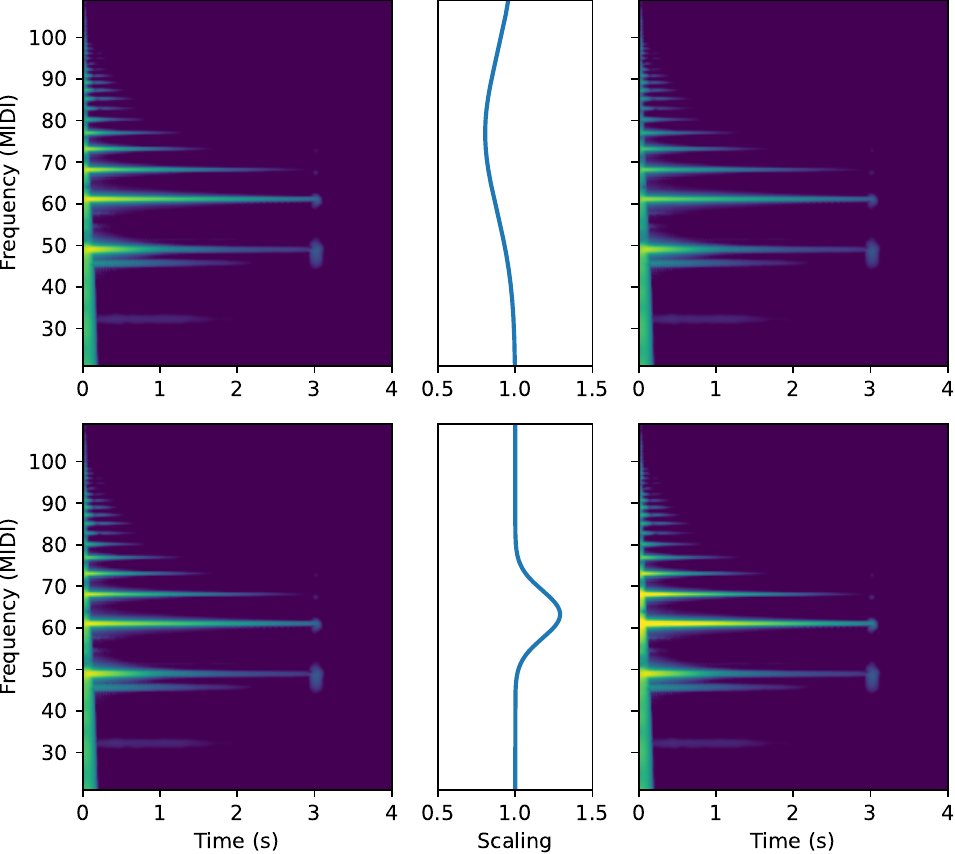}
\caption{Examples (by row) of sampled Gaussian equalization curves (center) applied to CQT spectrograms (left) to produce equalized spectrograms (right) for the timbre-invariance objective.}
\label{fig:equalization}
\end{figure}

\begin{figure}[t]
\centering
\includegraphics[width=0.99\columnwidth]{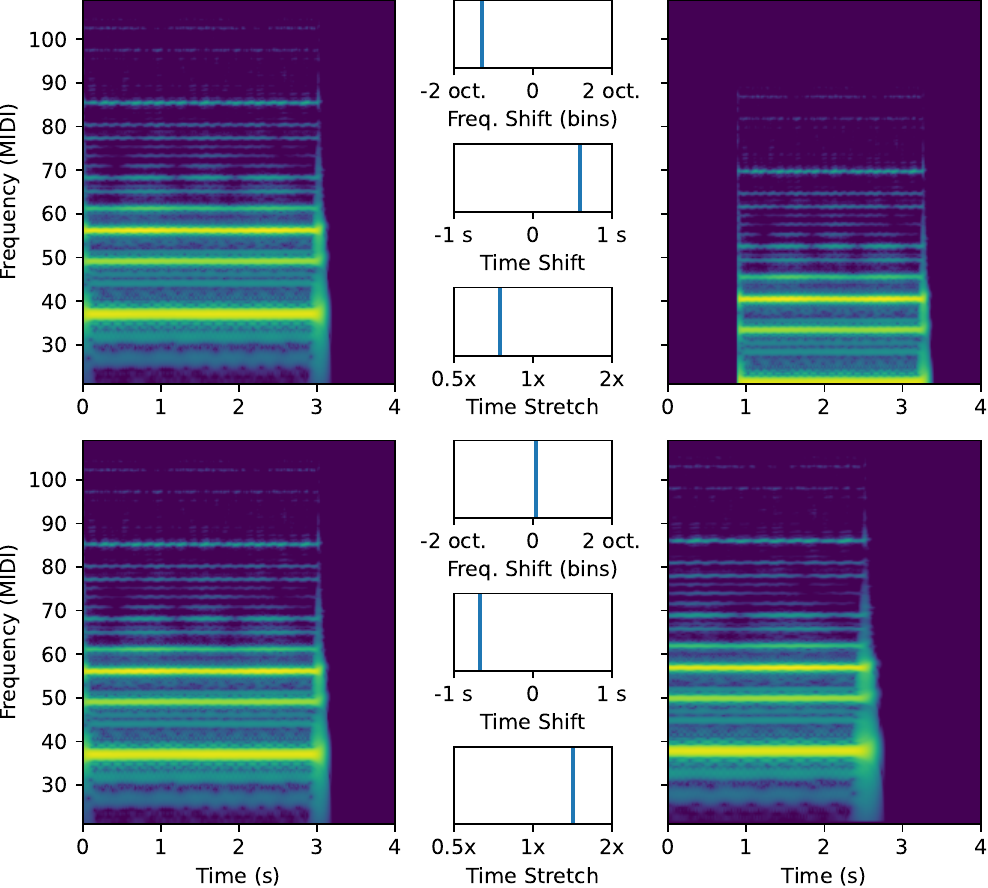}
\caption{Examples (by row) of sampled geometric transformations (center) applied to CQT spectrograms (left) to produce transformed spectrograms (right) for the geometric-equivariance objective.}
\label{fig:geometric}
\end{figure}

\subsection{Timbre Invariance}\label{sec:timbre_invariance}
The self-supervised objectives outlined in Section \ref{sec:energy_concentration} provide a reasonable foundation for estimating multi-pitch salience, but they do not yield particularly confident or separable output, and distribute energy broadly across harmonics.
Moreover, they are heavily influenced by the timbral properties, \textit{i.e.}, relative harmonic amplitudes, of the input audio.
In order to alleviate these issues, we introduce a \textit{timbre-invariance} objective, motivated by the notion that the same pitch content can be produced by many different musical instruments.
As such, we perform timbre transformations by applying Gaussian equalization curves $EQ(\cdot, \mu, \sigma, A)$
%\begin{equation}
%\label{eq:equalization}
%EQ(f, \mu, \sigma, A) = 1 + (1 - A) * e^{-\frac{1}{2} * \left( \frac{f - \mu}{\sigma} \right)^2}
%\end{equation}
to scale the input HCQT features, inspired by previous data augmentation strategies for MIR \cite{schluter2015exploring, abesser2021jazz}.
Gaussian parameters $\mu \in [0, K - 1]$, $\sigma \in [0, 2b_{oct}]$, and $A \in [0.625, 1.375]$ are uniformly sampled for each audio signal.
Equalization curves are applied across each channel of $X_\mathcal{H}$ by performing nearest-neighbor interpolation with respect to the scaling factor computed for frequency bins aligned with the first harmonic, producing equalized features $X_{\mathcal{H}}^{(EQ)}$, which are clipped to the range $[0, 1]$. An illustration of various sampled equalization curves and their application to an example CQT spectrogram is presented in Figure \ref{fig:equalization}.
Given model $\mathcal{F}(\cdot)$ and a set of equalized HCQT features $X_{\mathcal{H}}^{(EQ)}$, a timbre-invariance loss is then computed as
\begin{equation}
\label{eq:timbre_loss}
\mathcal{L}_{tmb} = \frac{1}{N} \sum^{N - 1}_{n = 0} \sum^{K - 1}_{k = 0}
\mathcal{B}\hspace{-0.15em}\left( \mathcal{F}\hspace{-0.15em}\left( X_{\mathcal{H}}^{(EQ)} \right)\hspace{-0.25em}[k, n], \hat{Y}[k, n] \right),
\end{equation}
where $\mathcal{B}(\cdot, \cdot)$ is the standard BCE loss, computed over the multi-pitch salience-gram estimated from the equalized features with respect to the salience-gram produced for the original features.
Applying random equalizations will inevitably boost and cut the various harmonics within each signal.
In conjunction with the objectives from Section \ref{sec:energy_concentration} this can allow the model to discover more efficient representations, \textit{i.e.}, the corresponding implicit multi-pitch salience-grams.

\subsection{Geometric Equivariance}\label{sec:geometric_invariance}
Our final self-supervised objective is based around the geometric-equivariance properties of the CQT noted in Section \ref{sec:feature_extraction}, which have been exploited in previous works on SSL for monophonic pitch estimation \cite{gfeller2020spice, riou2023pesto}.
As the input HCQT features are equivariant to translations in time and frequency, as well as distortions in time, we adopt a \textit{geometric-equivariance} objective to ensure the output representation shares the same equivariance properties.
We perform geometric transformations $GM(\cdot, \Delta_{k}, \Delta_{n}, \gamma)$ to shift and stretch the input HCQT features accordingly.
Integer transformation parameters $\Delta_{k} \in [-2b_{oct}, 2b_{oct}]$, $\Delta_{n} \in [-\frac{N}{4}, \frac{N}{4}]$ ($\frac{N}{4} \approx 1$ second for our experiments) are sampled uniformly, and real-valued parameter $\gamma \in [0.5, 2]$ is randomly sampled using two uniform distributions ($[0.5,1]$ and $[1,2]$) with equal weights.
The transformations are applied to all harmonic channels to produce features $X_{\mathcal{H}}^{(GM)}$, filling $0$ in for unknown values. An illustration of various sampled transformation parameters and their application to an example CQT spectrogram is presented in Figure \ref{fig:geometric}.
Given model $\mathcal{F}(\cdot)$ and a set of transformed HCQT features $X_{\mathcal{H}}^{(GM)}$, a geometric-equivariance loss is then computed as
\begin{equation}
\label{eq:geometric_loss}
\mathcal{L}_{geo} = \frac{1}{N} \sum^{N - 1}_{n = 0} \sum^{K - 1}_{k = 0}
\mathcal{B}\hspace{-0.15em}\left( \mathcal{F}\hspace{-0.15em}\left( X_{\mathcal{H}}^{(GM)} \right)\hspace{-0.25em}[k, n], \hat{Y}[k, n]^{(GM)} \right),
\end{equation}
where $\hat{Y}^{(GM)}$ is the multi-pitch salience-gram produced for the original features transformed in accordance with $X_{\mathcal{H}}^{(GM)}$.
Apart from inducing properties of geometric equivariance, this objective can also encourage the allocation of energy to the F0.
Specifically, certain transformations effectively low-pass filter the input features, discarding higher-order harmonics, while others effectively high-pass filter the input features, discouraging energy at higher harmonics if the output of the model adequately represents pitch salience.

\section{Experiments}\label{sec:experiments}
\begin{table*}[ht]
\footnotesize
\begin{center}
  \setlength{\tabcolsep}{0.575em}
  \renewcommand{\arraystretch}{1.25}
  \begin{tabular}{|c||c|c|c||c|c|c||c|c|c||c|c|c||c|c|c|}
    \hline
    & \multicolumn{3}{c||}{\textbf{Bach10}} & \multicolumn{3}{c||}{\textbf{URMP}} & \multicolumn{3}{c||}{\textbf{Su}} & \multicolumn{3}{c||}{\textbf{TRIOS}} & \multicolumn{3}{c|}{\textbf{GuitarSet}} \\
    \hline
    \textbf{Method} & $\mathit{P}$ & $\mathit{R}$ & $\mathit{Acc}$ & $\mathit{P}$ & $\mathit{R}$ & $\mathit{Acc}$ & $\mathit{P}$ & $\mathit{R}$ & $\mathit{Acc}$ & $\mathit{P}$ & $\mathit{R}$ & $\mathit{Acc}$ & $\mathit{P}$ & $\mathit{R}$ & $\mathit{Acc}$ \\
    \hline
    \hline
    Deep-Salience & $86.0$ & $61.0$ & $55.5$ & $91.1$ & $72.8$ & $67.6$ & $74.1$ & $47.9$ & $40.4$ & $92.3$ & $39.4$ & $38.1$ & $77.7$ & $70.6$ & $57.7$ \\
    \hline
    Basic-Pitch & $90.2$ & $75.7$ & $70.0$ & $88.2$ & $71.1$ & $64.9$ & $54.2$ & $44.1$ & $32.8$ & $88.2$ & $44.2$ & $41.5$ & \textcolor{gray}{$80.9$} & \textcolor{gray}{$75.9$} & \textcolor{gray}{$64.0$} \\
    \hline
    Timbre-Trap & $81.2$ & $84.2$ & $70.4$ & \textcolor{gray}{$80.5$} & \textcolor{gray}{$88.8$} & \textcolor{gray}{$73.3$} & $52.1$ & $53.0$ & $34.8$ & $69.4$ & $49.7$ & $40.4$ & $48.6$ & $75.6$ & $41.3$ \\
    \hline
    \hline
    CREPE$^{(multi)}$ & \textcolor{gray}{$82.0$} & \textcolor{gray}{$24.8$} & \textcolor{gray}{$23.5$} & $88.2$ & $37.3$ & $35.6$ & $63.9$ & $20.4$ & $17.9$ & $87.6$ & $20.3$ & $19.6$ & $70.7$ & $53.9$ & $43.9$ \\
    \hline
    PESTO$^{(multi)}$ & $81.7$ & $23.0$ & $21.9$ & $65.7$ & $27.8$ & $23.9$ & $69.9$ & $21.9$ & $19.6$ & $78.4$ & $18.4$ & $17.3$ & $61.3$ & $50.7$ & $36.5$ \\
    \hline
    \hline
    SS-MPE ($\mathcal{L}$) & $65.1$ & $82.5$ & $57.1$ & $61.6$ & $63.9$ & $44.4$ & $50.4$ & $52.8$ & $34.1$ & $65.8$ & $51.1$ & $38.9$ & $49.9$ & $67.5$ & $39.2$ \\
    \hline
    SS-MPE ($\mathcal{L}_t$) & $65.4$ & $66.5$ & $49.1$ & $66.0$ & $49.5$ & $37.3$ & $52.3$ & $45.4$ & $31.3$ & $65.3$ & $43.7$ & $34.2$ & $54.4$ & $55.4$ & $36.4$ \\
    \hline
    SS-MPE ($\mathcal{L}_g$) & $63.0$ & $81.4$ & $55.0$ & $63.8$ & $48.9$ & $36.4$ & $54.4$ & $38.8$ & $28.6$ & $70.3$ & $39.1$ & $31.3$ & $59.5$ & $58.1$ & $40.2$ \\
    \hline
    SS-MPE ($\mathcal{L}_e$) & $59.9$ & $78.6$ & $51.4$ & $58.8$ & $49.2$ & $34.5$ & $50.7$ & $39.5$ & $27.7$ & $63.9$ & $40.6$ & $30.3$ & $52.0$ & $60.4$ & $37.0$ \\
    \hline
  \end{tabular}
\end{center}
\caption{Comparison of precision (\textit{P}), recall (\textit{R}), and accuracy (\textit{Acc}) scores (in percentage points) on MPE and AMT datasets for our self-supervised framework (SS-MPE). Results are also reported for supervised MPE methods Deep-Salience \cite{bittner2017deep}, Basic-Pitch \cite{bittner2022lightweight}, and Timbre-Trap \cite{cwitkowitz2024timbre}, as well as multi-pitch estimates derived from the salience-grams of monophonic models CREPE \cite{kim2018crepe} and PESTO \cite{riou2023pesto}. Grayed values indicate data was utilized during training.}
\label{tab:polyphonic_results}
\end{table*}

In order to demonstrate the efficacy of SS-MPE, we design experiments where the model is trained on audio from the NSynth \cite{engel2017neural} dataset and evaluated on various MPE datasets.
Our framework is trained with all the objectives defined in Sections \ref{sec:energy_concentration}-\ref{sec:geometric_invariance}, \textit{i.e.}, with overall loss
\begin{equation}
\label{eq:total_loss}
\mathcal{L} = \mathcal{L}_{har} + \mathcal{L}_{sup} + \mathcal{L}_{spr} + \mathcal{L}_{tmb} + \mathcal{L}_{geo}.
\end{equation}
However, we also experiment with various ablations to the training objective in order to observe the utility of timbre-invariance and geometric-equivariance.
In particular, we also train the framework using the following loss variations:
\begin{gather}
\label{eq:energy_ablation}
\mathcal{L}_e = \mathcal{L}_{har} + \mathcal{L}_{sup} + \mathcal{L}_{spr}, \\
\label{eq:timbre_ablation}
\mathcal{L}_t = \mathcal{L}_e + \mathcal{L}_{tmb}, \\
\label{eq:geometric_ablation}
\mathcal{L}_g = \mathcal{L}_e + \mathcal{L}_{geo}.
\end{gather}
These ablations correspond to experiments with the energy concentration objectives only, the energy concentration objectives with the timbre-invariance objective, and the energy concentration objectives with the geometric-invariance objective.
Each experiment was executed on a single GeForce RTX 2080 TI graphics card and took between 3 to 4 days.

\subsection{Datasets}\label{sec:datasets}
\textbf{NSynth} is a collection of over 300k music note samples synthesized from commercial libraries \cite{engel2017neural}.
It covers $11$ instrument classes, $1006$ unique instruments, pitches spanning the $88$ keys of a standard MIDI piano, and several note velocities.
Each note sample corresponds to $4$ seconds of audio, with the note onset and release (if applicable) occurring at $t = 0$ and $t = 3$, respectively.
An example of a note sample from NSynth is presented in Figure \ref{fig:sample}.
All audio within NSynth is monophonic, and any note samples with pitch outside of the supported range of our HCQT module are discarded.
We follow the provided splits for training and validation, but use only a subset of $200$ samples for validation to reduce overall validation time.

Evaluation is conducted on a variety of MPE and AMT datasets, mostly comprising classical music. We evaluate on the full datasets, using multitrack mixtures if applicable.

\textbf{Bach10} contains multitrack audio along with multi-pitch annotations for $10$ short four-part Bach chorales featuring a violin, clarinet, saxophone, and bassoon \cite{duan2010multiple}.

\textbf{URMP} contains multitrack audio along with multi-pitch annotations for $44$ classical music pieces arranged for string, woodwind, and brass instruments spanning $14$ instrument classes \cite{li2018creating}.
The dataset features $11$ duets, $12$ trios, $14$ quartets, and $7$ quintets.

The \textbf{Su} dataset contains audio mixtures and note-level annotations for $3$ piano solo, $3$ string quartet, $2$ piano quintet, and $2$ violin sonata passages \cite{su2016escaping}.

\textbf{TRIOS} contains multitrack audio along with note-level annotations for $4$ classical music trio passages and $1$ jazz trio passage \cite{fritsch2012master}. The dataset includes piano, string, woodwind, and brass instruments, and the jazz track features percussive elements.

\textbf{GuitarSet} contains audio along with multi-pitch and note-level annotations for 360 short monophonic and polyphonic recordings of an acoustic guitar \cite{quingyang2018guitarset}.

\subsection{Training \& Evaluation Details}\label{sec:training_evaluation_details}
All training is conducted with a batch size of $20$ using AdamW optimizer \cite{loshchilov2019decoupled}.
We set an initial learning rate of $10^{-4}$ and halve the learning rate whenever the validation loss has not decreased for $0.5$ epochs.
We also perform gradient-norm clipping using a value of $10$.
In each experiment, the model is only trained for $3$ epochs, which is sufficient to reach convergence.
Validation is conducted every $250$ batches, and the checkpoint with the lowest validation loss across training is selected for evaluation.
We utilize the community-standard \texttt{mir\_eval} package \cite{raffel2014mir_eval} to compute precision (\textit{P}), recall (\textit{R}), and accuracy (\textit{Acc}) of multi-pitch estimates with respect to the ground-truth $Y$ for the tracks within the MPE and AMT datasets described in Section \ref{sec:datasets}.
Accuracy is computed as the ratio between the number of pitch estimates within $0.5$ semitones of matched ground-truth pitches (\textit{true positives}) to the total number of pitch estimates (\textit{true positives} and \textit{false positives}) plus the number of missed ground-truth pitch estimates (\textit{false negatives}).
Final results are computed by averaging across all tracks within an individual dataset.

\subsection{Baselines}\label{sec:baselines}
We compare our self-supervised framework (SS-MPE) to various supervised MPE baselines, including Deep-Salience \cite{bittner2017deep}, Basic-Pitch \cite{bittner2022lightweight}, and Timbre-Trap \cite{cwitkowitz2024timbre}.
Deep-Salience is a convolutional neural network (CNN) with an HCQT frontend that was trained on mixtures created from F0-annotated stems of MedleyDB \cite{bittner2014medleydb}.
Basic-Pitch extends this methodology and incorporates ideas from the Onsets \& Frames \cite{hawthorne2018onsets} model to generate multi-pitch and note predictions using a single architecture. It is trained on a variety of AMT and MPE datasets including Slakh \cite{manilow2019cutting}, MAESTRO \cite{hawthorne2018enabling}, MedleyDB stems, GuitarSet \cite{quingyang2018guitarset}, and vocal data.
Timbre-Trap leverages an audio reconstruction objective to more efficiently learn from a portion of URMP \cite{li2018creating} (see Section \ref{sec:model} for more details).
We also obtain multi-pitch estimates from the pitch salience-grams generated by monophonic pitch estimation models CREPE \cite{kim2018crepe} and PESTO \cite{riou2023pesto}.
CREPE is a supervised CNN trained on MIR-1K \cite{hsu2010improvement}, Bach10 stems, RWC-Synth \cite{mauch2014pyin}, MedleyDB stems, MDB-STEM-Synth \cite{salamon2017analysis}, and NSynth.
PESTO is a self-supervised framework featuring a CNN trained on MIR-1K.
Although these are monophonic pitch estimation methods, we include them because our framework is also trained on monophonic audio.
For all baselines, frame-level F0 estimates are obtained by performing local peak-picking across frequency and applying a threshold (0.5 for Timbre-Trap and 0.3 for all others).

\begin{figure*}[t]
\centering
\includegraphics[width=0.99\linewidth]{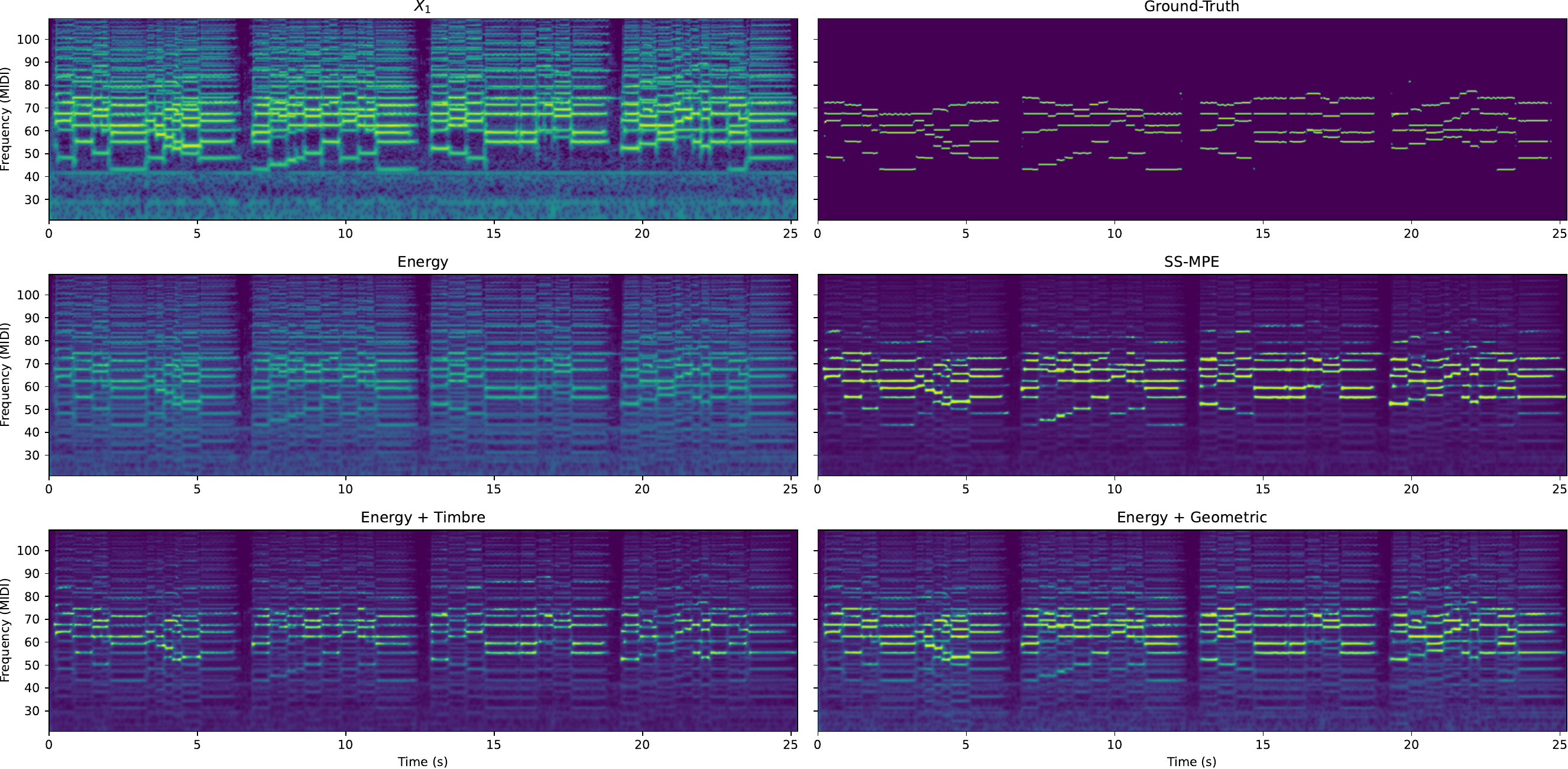}
\caption{CQT spectrogram $X_1$ for track \texttt{01-AchGottundHerr} of Bach10 \cite{duan2010multiple} along with the multi-pitch salience-gram output generated using our full self-supervised framework (SS-MPE) and the output for the framework with training objective ablations.}
\label{fig:inference}
\end{figure*}

\subsection{Results \& Discussion}\label{sec:results_and_discussion}

Results for all experiments and baselines are presented in Table \ref{tab:polyphonic_results}, and an example of the multi-pitch salience-gram output for SS-MPE and each ablation is provided in Figure \ref{fig:inference}.
Generally, we observe that the performance of SS-MPE approaches and in some cases rivals supervised MPE methods, while greatly outperforming the simple salience thresholding scheme applied to the monophonic baselines for most datasets.
However, we do note that the monophonic baselines perform comparably to SS-MPE on GuitarSet, roughly half of which consists of monophonic audio.

These promising results are achieved without applying any weighting to the objectives, performing any tuning on the threshold for positive predictions, or calibrating the trained model with annotated data as in \cite{gfeller2020spice, riou2023pesto}.
Moreover, our framework is trained entirely on 4-second monophonic note samples from NSynth \cite{engel2017neural}, yet remarkably it generalizes to sequences of polyphonic audio with pitch fluctuations such as vibrato.
This is not true for the monophonic baselines, which have trouble identifying more than one pitch at a time, as evidenced by their high precision and low recall scores.

The full framework with all objectives ($\mathcal{L}$) achieves the best overall performance compared to the ablations and self-supervised PESTO.
It also exhibits the most sparsity when predicting multi-pitch salience-grams.
However, the timbre-invariance and geometric-invariance objectives applied without one another ($\mathcal{L}_t$ and $\mathcal{L}_g$) still generally contribute to improved performance over the energy-only objectives ($\mathcal{L}_e$).
These experiments also suggest the contribution of timbre-invariance and geometric-invariance is varied on specific datasets. For instance, geometric-equivariance in isolation ($\mathcal{L}_g$) improves over timbre-invariance ($\mathcal{L}_t$) in isolation on Bach10 and GuitarSet, whereas the opposite is true for URMP, Su, and TRIOS.
We also speculate that NSynth may already intrinsically encode geometric-equivariance with respect to pitch to some degree, since it comprises notes played over the entire pitch range of each instrument.

While the presented results are extremely encouraging, there is still plenty of room for improvement.
In particular, the current framework is relatively susceptible to false-alarm predictions, especially those caused by harmonics and artifacts (\textit{e.g.}, percussive note attack or release).
Additionally, sometimes the F0 is missed entirely, potentially due to the lack of realistic training data.
Future work will consist of improving the existing formulation, incorporating more augmentation techniques for increased robustness (\textit{e.g.}, noise-invariance to reduce the effect of artifacts or time-varying transformations to simulate pitch fluctuations in real recordings), and scaling the framework to polyphonic music datasets.

\section{Conclusion}\label{sec:conclusion}
In this paper, we have introduced a fully self-supervised framework to generate multi-pitch estimates.
Our framework leverages a harmonic spectrogram, an autoencoder architecture, and self-supervised objectives designed to concentrate sparse energy around strong fundamental frequency candidates and exploit the properties of timbre invariance and geometric equivariance.
The model is trained entirely using isolated synthetic single-note samples, yet can generalize to conventional polyphonic data.
No fine-tuning is required within our framework, and performance approaches that of supervised methods.
Future work will consist of improving and extending the current methodology, as well as scaling it to larger and more comprehensive audio datasets.

\onecolumn
\begin{multicols}{2}
\section*{Acknowledgements}
This work is partially supported by National Science Foundation (NSF) grants No. 1846184 and 2222129, and synergistic activities funded by NSF grant DGE-1922591.

\bibliography{ssmpe}
\bibliographystyle{icml2024}
\end{multicols}

\appendix
\onecolumn
\section{Continuous Binary Cross-Entropy}
\begin{figure}[h]
\centering
\includegraphics[width=0.98\columnwidth]{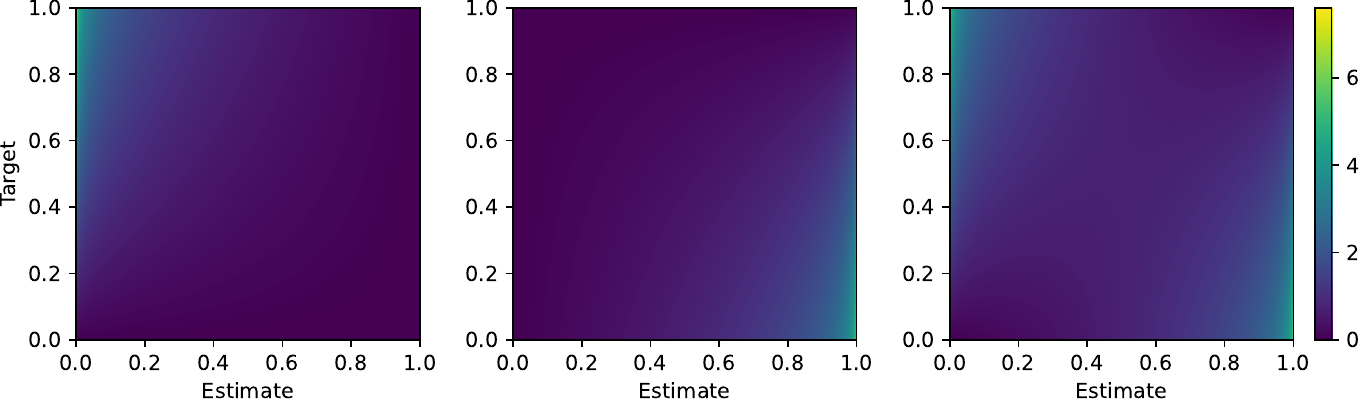}
\caption{Positive activation (left), negative activation (center), and standard (right) BCE loss over continuous-valued targets and estimates.}
\label{fig:bce}
\end{figure}

\section{Validation Loss Curves}
\begin{figure}[h]
\centering
\includegraphics[width=0.98\columnwidth]{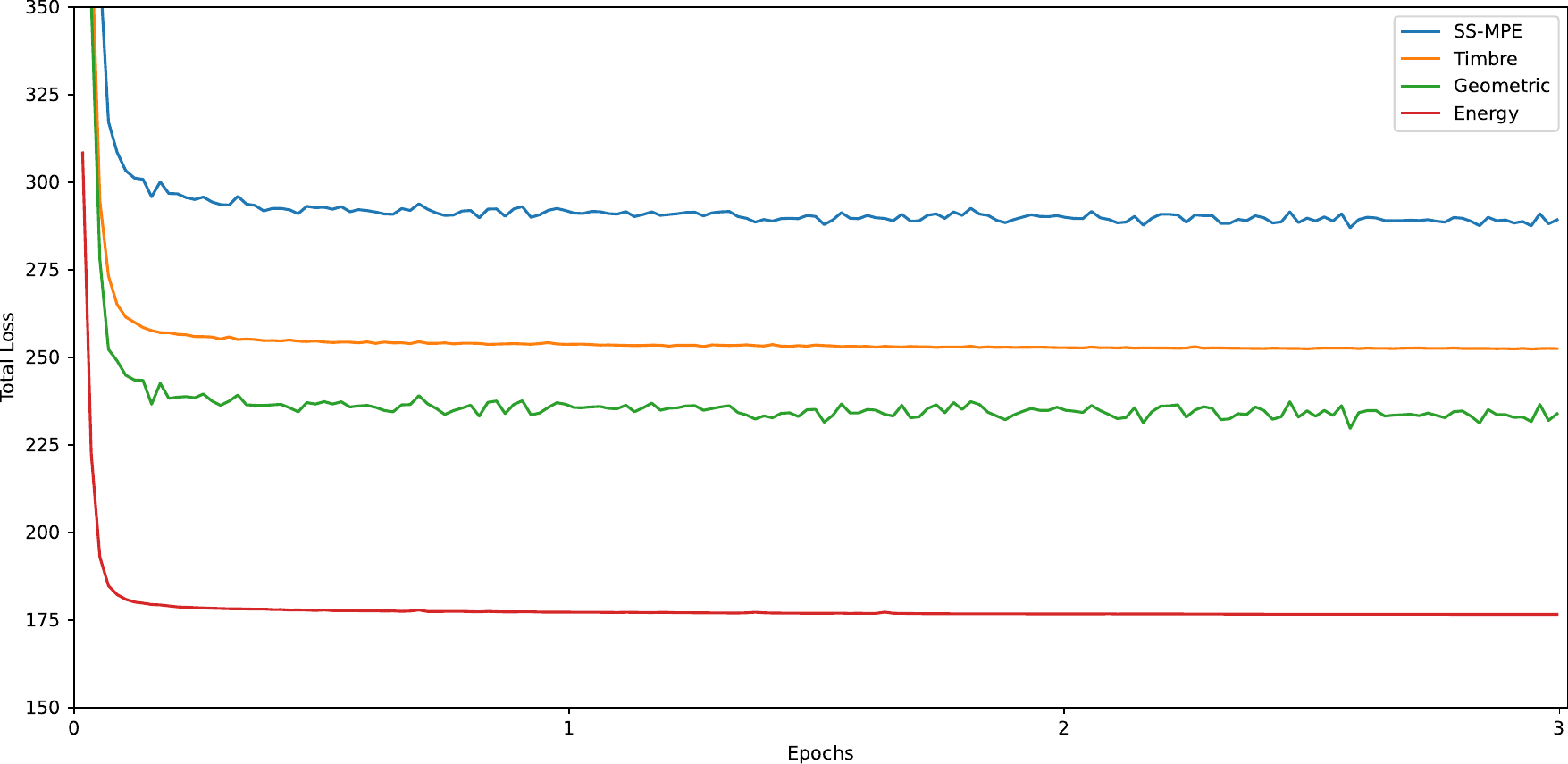}
\caption{Validation loss computed at checkpoints of SS-MPE and each ablation.}
\label{fig:loss}
\end{figure} 

\newpage
\section{Additional Inference Examples}
\begin{figure}[h]
\centering
\includegraphics[width=0.9\linewidth]{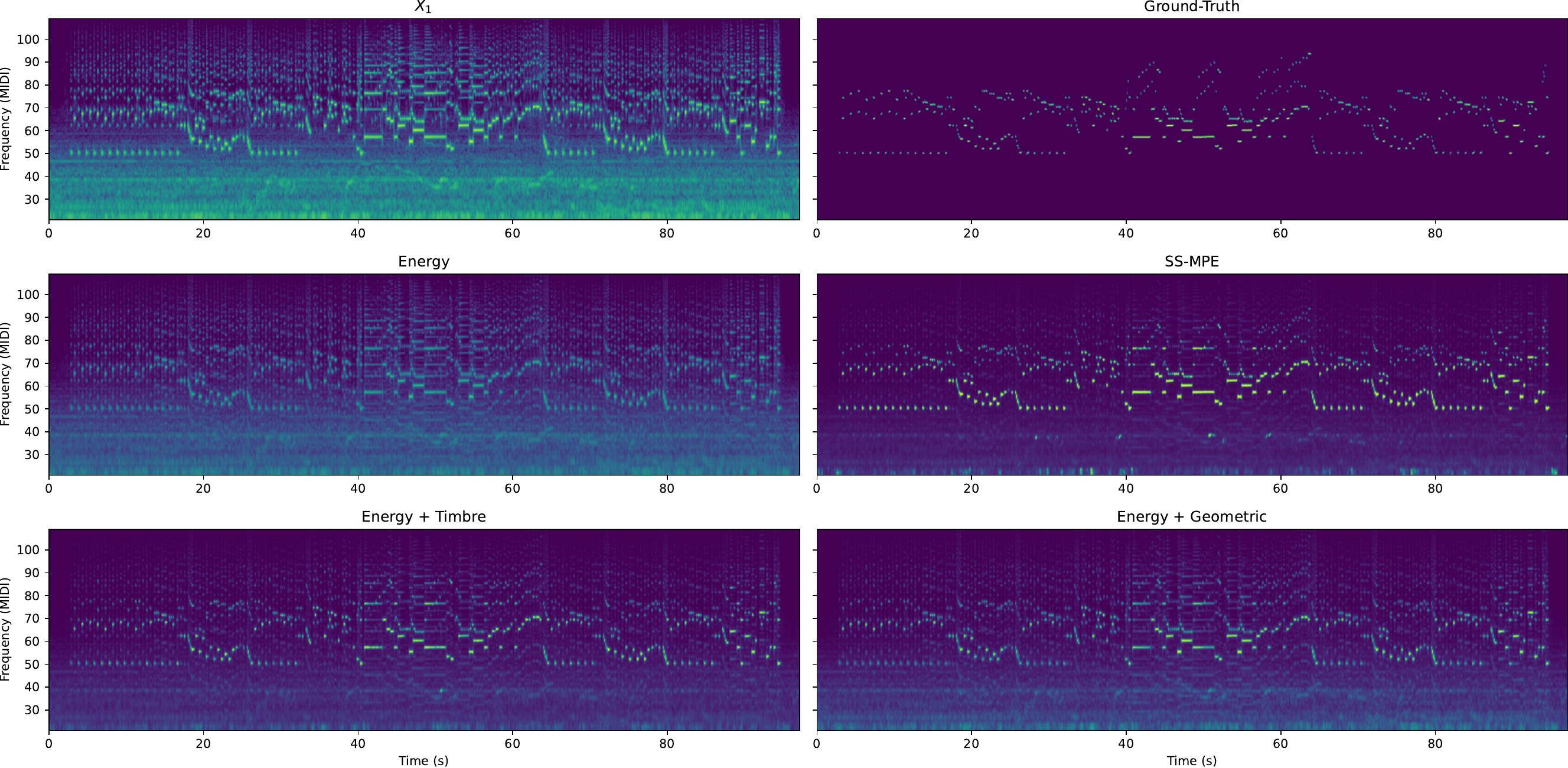}
\caption{Inference for track \texttt{03\_Dance\_fl\_cl} of URMP \cite{li2018creating}.}
\label{fig:inference_urmp}
\end{figure}

\begin{figure}[h]
\centering
\includegraphics[width=0.9\linewidth]{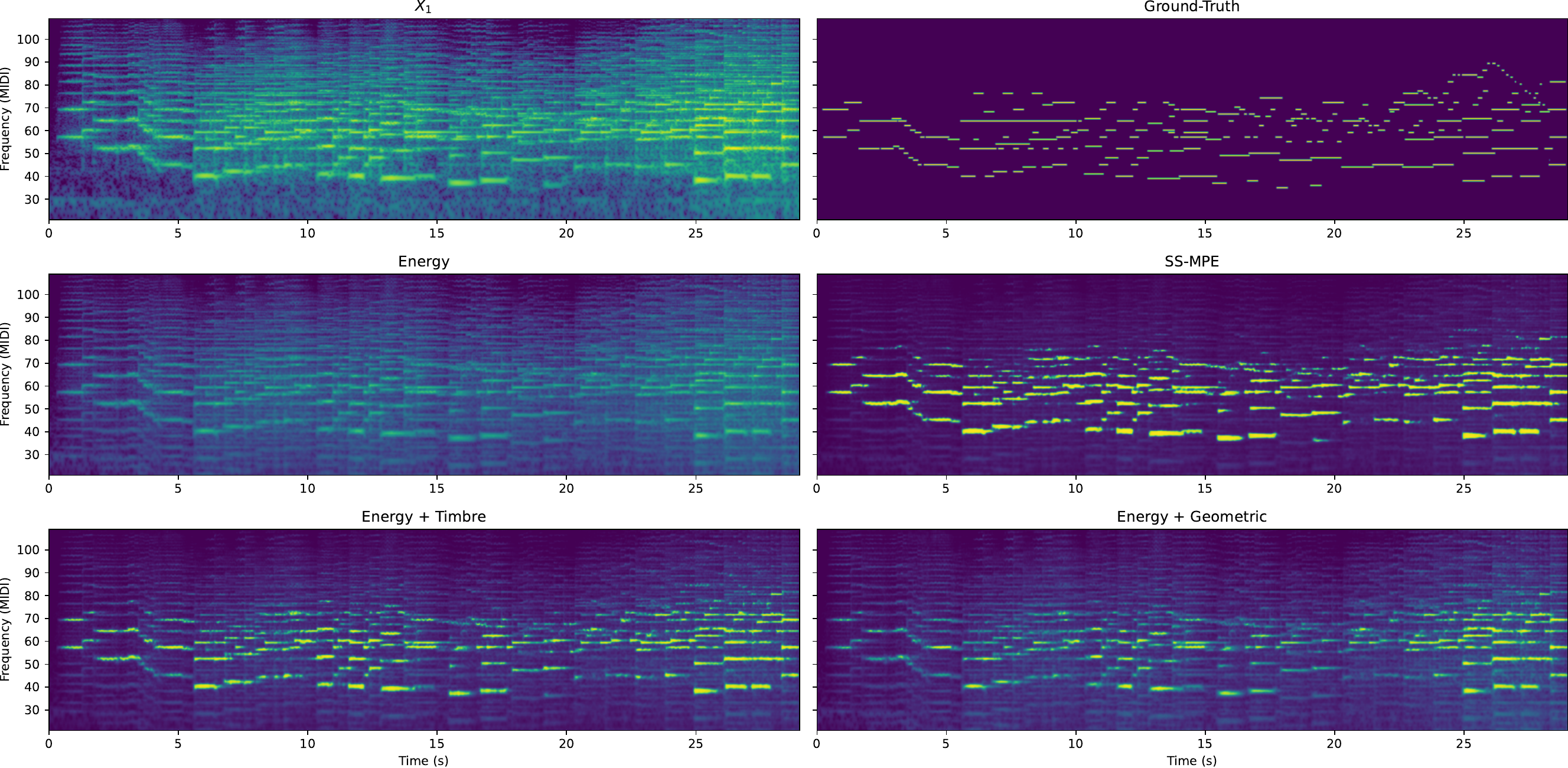}
\caption{Inference for track \texttt{PQ03\_Farranc} of Su \cite{su2016escaping}.}
\label{fig:inference_su}
\end{figure}

\begin{figure}[h]
\centering
\includegraphics[width=0.9\linewidth]{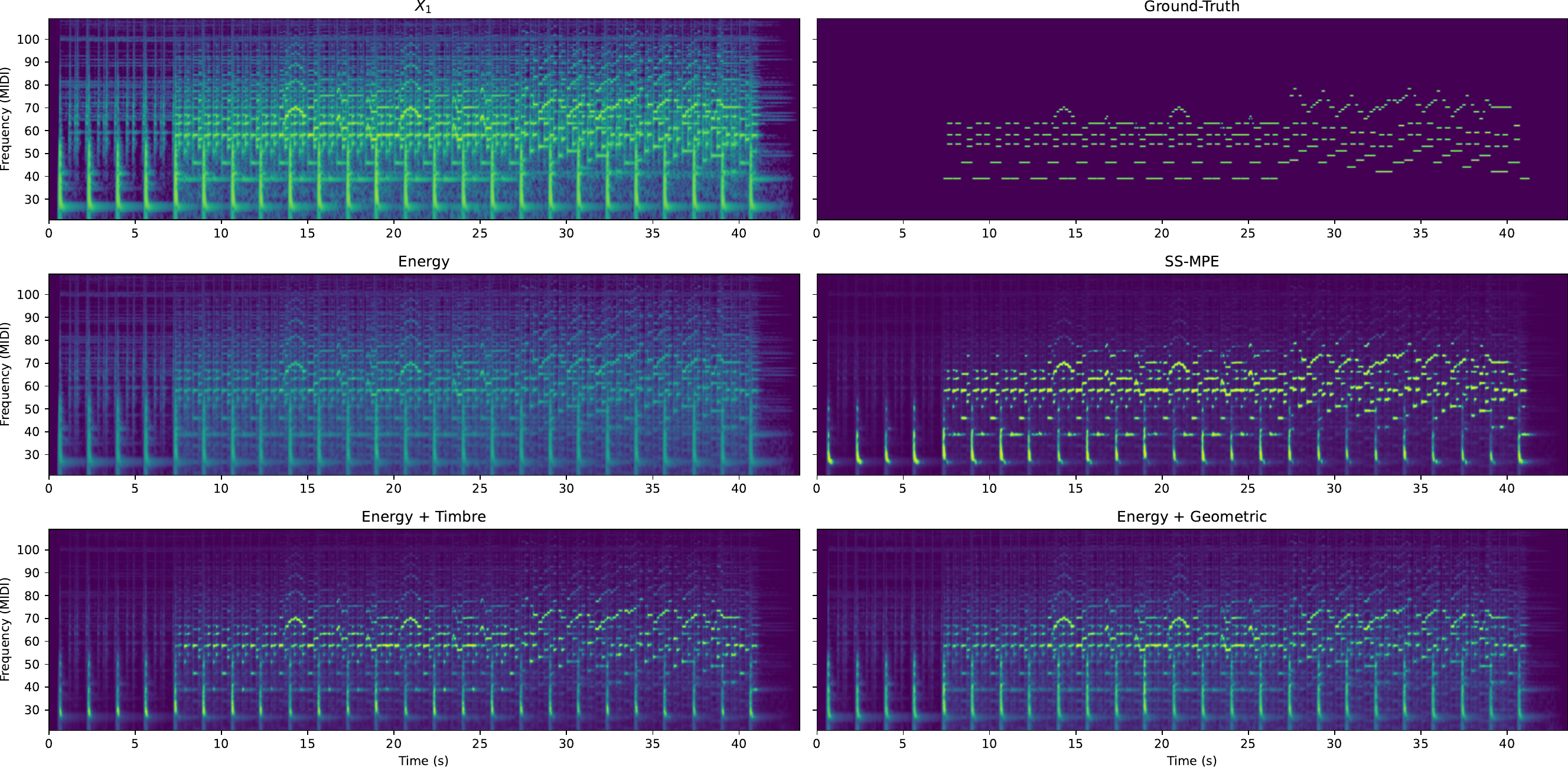}
\caption{Inference for track \texttt{take\_five} of TRIOS \cite{fritsch2012master}.}
\label{fig:inference_trios}
\end{figure}

\begin{figure}[h]
\centering
\includegraphics[width=0.9\linewidth]{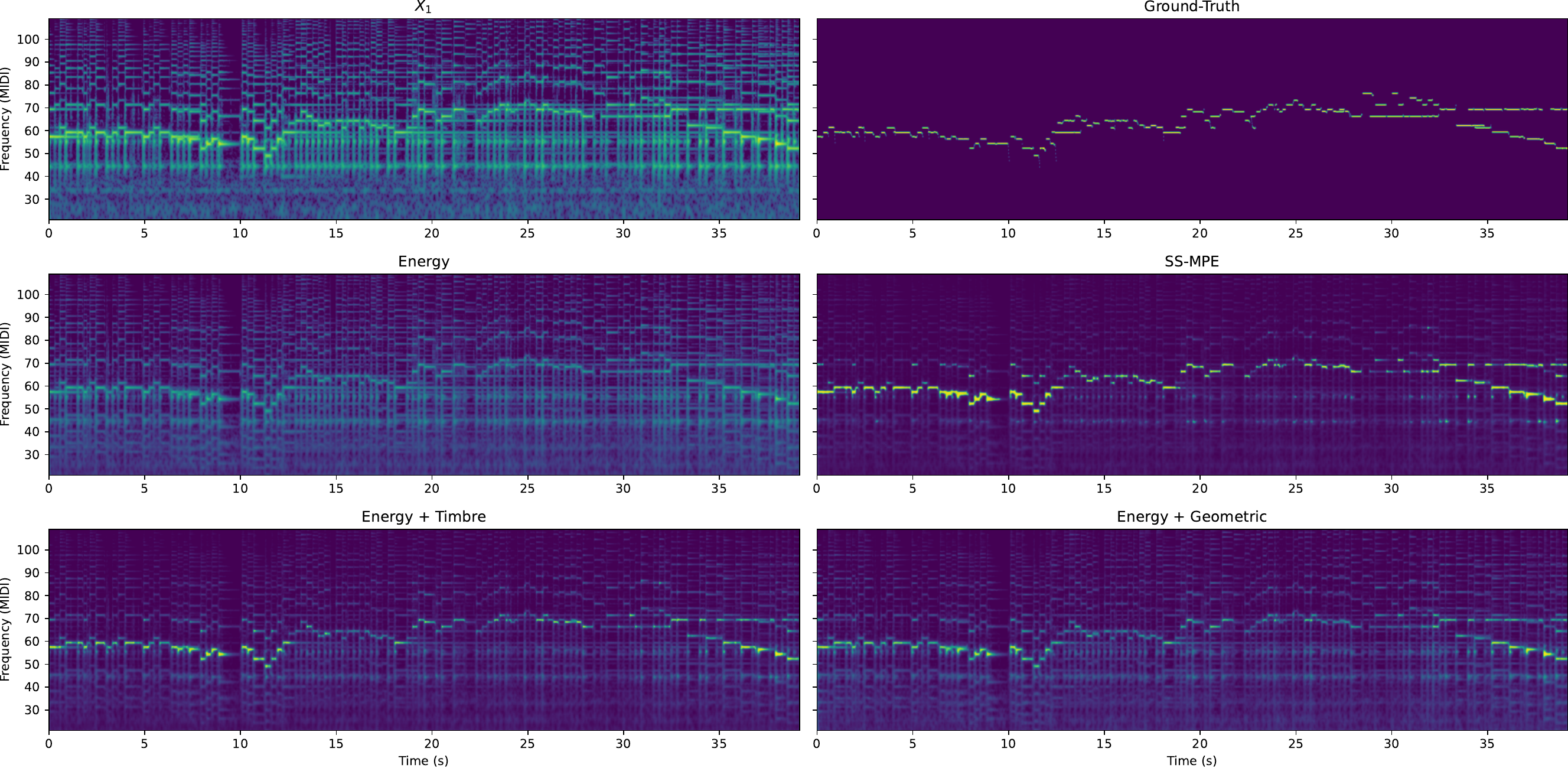}
\caption{Inference for track \texttt{00\_Funk3-98-A\_solo} of GuitarSet \cite{quingyang2018guitarset}.}
\label{fig:inference_guitarset}
\end{figure}

\end{document}